\documentclass[seceq]{ptptex}
\usepackage{graphicx}
\usepackage{dcolumn}
\usepackage{bm}
\usepackage{longtable}

\newlength{\Tatescale}
\newlength{\figwidth}
\newcommand{\Cut}[1]{}
 
\newcommand{\beq}{\begin{eqnarray}}
\newcommand{\eeq}{\end{eqnarray}}

\newcommand{\bq}{\mbox{\boldmath $q$}}
\newcommand{\bp}{\mbox{\boldmath $p$}}
\newcommand{\bk}{\mbox{\boldmath $k$}}
\newcommand{\br}{\mbox{\boldmath $r$}}
\newcommand{\bx}{\mbox{\boldmath $x$}}
\newcommand{\by}{\mbox{\boldmath $y$}}
\newcommand{\bz}{\mbox{\boldmath $z$}}

\newcommand{\bw}{\mbox{\boldmath $w$}}
\newcommand{\bL}{\mbox{\boldmath $L$}}

\newcommand{\bP}{\mbox{\boldmath $P$}}

\newcommand{\bS}{\mbox{\boldmath $S$}}

\newcommand{\bsigma}{\mbox{\boldmath $\sigma$}}

\newcommand{\brs}{\mbox{\scriptsize \boldmath $r$}}
\newcommand{\bxs}{\mbox{\scriptsize \boldmath $x$}}
\newcommand{\bys}{\mbox{\scriptsize \boldmath $y$}}
\newcommand{\bps}{\mbox{\scriptsize \boldmath $p$}}
\newcommand{\bqs}{\mbox{\scriptsize \boldmath $q$}}
\newcommand{\bks}{\mbox{\scriptsize \boldmath $k$}}

\newcommand{\pslash}{p\kern-1ex /}
\newcommand{\kslash}{k\kern-1ex /}
\newcommand{\qslash}{q\kern-1ex /}
\newcommand{\lslash}{l\kern-1ex /}
\newcommand{\sslash}{s\kern-1ex /}
\newcommand{\paslash}{p_a\kern-2ex /}
\newcommand{\pbslash}{p_b\kern-2ex /}
\newcommand{\Dslash}{{\cal D}\kern-1.5ex /}
\newcommand{\dslash}{\partial\kern-1.2ex /}

\title{
 Extraction of Hadron Interactions above Inelastic Threshold in Lattice QCD %
}

\author{
Sinya \textsc{Aoki}$^{1,2}$\footnote{corresponding author: saoki@het.ph.tsukuba.ac.jp}, 
Noriyoshi \textsc{Ishii}$^1$,
Takumi \textsc{Doi}$^{1}$,
Tetsuo \textsc{Hatsuda}$^{3,4,5}$, 
Yoichi \textsc{Ikeda}$^{6}$,
Takashi \textsc{Inoue}$^{7}$
Keiko \textsc{Murano}$^{5}$,
Hidekatsu \textsc{Nemura}$^{8}$,  
Kenji \textsc{Sasaki}$^{1}$\\
(HAL QCD Collaboration)\\
}
\inst{
$^1$ Graduate School of Pure and Applied Sciences, University of Tsukuba,\\ Tsukuba 305-8571, Japan\\
$^2$Center for Computational Sciences, University of Tsukuba,\\ Tsukuba 305-8577, Japan\\
$^3$Department of Physics, The University of Tokyo,  Tokyo 113-0033, Japan\\
$^4$IPMU, The University of Tokyo, Kashiwa 277-8583, Japan\\
$^5$Theoretical Research Division, Nishina Center, RIKEN, Wako 351-0198, Japan\\
$^6$Department of Physics, Tokyo Institute of Technology,  Tokyo 152-8551, Japan\\
$^7$Nihon University, College of Bioresource Sciences,  Fujisawa 252-0880, Japan\\
$^8$Department of Physics, Tohoku University, Sendai 980-8578, Japan
}

\abst{  We propose a new method to extract hadron interactions above inelastic threshold  
  from  the Nambu-Bethe-Salpter amplitude  in lattice QCD.  We consider the scattering such as
  $A+B\rightarrow  C+D$, where $A,B,C,D$ are names of different 1-particle states. 
An extension to cases where particle productions occur during scatterings is also discussed.\\
\underline{keywords}: nuclear potential, hadron interaction, lattice QCD, inelastic scattering
}

\PTPindex{164,232,234}  

\begin{document}

\maketitle

\section{Introduction}
\label{sec:intro}
 
 The origin of the nuclear force is one of the 
 major unsolved  problems in particle and nuclear physics
  even after the establishment of the quantum chromodynamics (QCD).
  Recently, three of the present authors proposed a new
  approach to extract the NN interactions below inelastic threshold 
  in lattice QCD \cite{Ishii:2006ec,Aoki:2008hh,Aoki:2009ji}.  
  Through the Nambu-Bethe-Salpeter (NBS) wave function,     
  the energy-independent  but non-local potential $U(\br, \br')$ is so defined
  that  the NBS wave function  obeys the 
  Schr\"{o}dinger type equation in finite volume.
  Since $U(\br, \br')$ is localized in its spatial coordinates due to confinement
   of quarks and gluons, the potential receives
   finite volume effect only weakly in a large box. Therefore, 
   once $U$ is determined and is appropriately extrapolated to 
   $L \rightarrow \infty$, one may simply use the Schr\"{o}dinger
   equation in the infinite space to calculate the scattering phase shifts
    and bound state spectra, which can be compared  with experimental data.    
    With this approach  we successfully extract not only the nucleon potential\cite{Ishii:2006ec,Aoki:2008hh,Aoki:2009ji,Ishii:2009zr} but also the hyperon potential\cite{Nemura:2008sp,Nemura:2009kc} below inelastic threshold in QCD.

Although this method is shown to be quite successful in order to describe elastic hadron interactions, the hadron interactions generally leads to inelastic scatterings as
the total energy of the system increases.           
In this paper, we extend our method to such inelastic scatterings in order to extract hadron interactions in general.
 In Sec.~\ref{sec:elastic}, we briefly summarize our method previously used to extract the potential in the elastic scattering. 
  In Sec.~\ref{sec:inelastic}, we present our main idea to analyze the inelastic scattering from lattice QCD in the finite volume. We here discuss the scattering such as $A+B\rightarrow C+D$ scattering, where $A,B,C,D$ represent some 1-particle states. This is a simplified version of the baryon scattering in the strangeness $S=-2$ and isospin $I=0$ channel, where $\Lambda\Lambda$, $N\Xi$ and $\Sigma\Sigma$ appear as asymptotic states if the total energy is larger than $2 m_\Sigma$.
In Sec.~\ref{sec:extension},
we discuss the extension of  our proposal to the scattering with particle productions such as $A+B\rightarrow A+B+C$. In Sec.~\ref{sec:summary} we summarize this paper  together with recent applications of our method in lattice QCD.
A preliminary account  of these results is given in Ref.\citen{Ishii:2011tq}.

\section{Hadron interactions below threshold: elastic scattering}
\label{sec:elastic}

In this section, we summarize our strategy, previously used to extract the potential between two hadrons below inelastic threshold in lattice QCD.

\subsection{Nambu-Bethe-Salpeter wave function}

A key quantity in our method is the equal-time Nambu-Bethe-Salpter (NBS) amplitude,
which we call  the ``NBS wave function"  throughout this paper.
 Let us consider the following NBS wave function for a particle $A$ and a particle $B$ in QCD
 with total energy $W$ in the center of mass system ({\it i.e.} the total three-momentum $\bP=0$) in the infinite box;
\begin{eqnarray}
\psi_{AB}^W(\br) e^{-Wt}
 = \lim_{\delta\rightarrow 0^+}
\langle 0 \vert  T\{\varphi_{A}(\by,t+\delta) \varphi_{B}(\bx,t)\} \vert AB; W,\bP=0 \rangle 
\label{eq:BS-def-def}
\end{eqnarray}   
where the relative coordinate is denoted as $\br = \bx - \by$. 
Here the   local operators for the $A$ and the $B$, which might be composite,
 are denoted by   $\varphi_{A}(\bx,t)$ and  $\varphi_{B}(\by,t)$
 with possible indices such as spinor or flavor being suppressed for simplicity.
  The QCD vacuum is denoted by $\vert 0 \rangle $, while the
 state $\vert AB; W,\bP=0  \rangle $ is a
 QCD eigenstate with the same quantum numbers as the $AB$ system.
 Note  that $\vert  AB; W,\bP=0  \rangle $ can be taken as
 a product of 1-particle asymptotic state $\vert A \rangle_{\rm in} \otimes \vert B \rangle_{\rm in}$
 in the infinite box, while this is not true in the finite box.
 
 If the total energy $W=E_k^A+E_k^B=\sqrt{m_A^2+\bk^2}+\sqrt{m_B^2+\bk^2}$ is smaller than the inelastic threshold $E_{th}$, $\psi_{AB}^W(\br)$ satisfies
 \begin{eqnarray}
\psi_{AB}^W(\br) = e^{i\bks \cdot\brs} &+& \int \frac{d^3p}{(2\pi)^3} \frac{e^{i\bp\cdot\br}}{(\bp^2-\bk^2-i\epsilon)} 
\frac{E_p^A+E_k^A}{4WE_p^A}T_{AB,AB}(p_A,p_B;k_A,k_B)  \nonumber \\
&+& {\cal I}(\br)
\end{eqnarray}
where $p_A=(E_p^A,\bp)$, $k_A=(E_k^A,\bk)$, $k_B=(E_k^B,-\bk)$
are on-shell 4 momenta, while $ p_B=(W-E_p^A,-\bp)$
is generally off-shell. Therefore $T_{AB,AB}(q_1,q_2,q_3,q_4)$ is in general off-shell T-matrix,
defined through the connected four-point Green's function $G_{AB,AB}^{(c)}(p_1,p_2;p_3,p_4)$
as
\begin{eqnarray}
G_{AB,AB}^{(c)}(p_1,p_2;p_3,p_4)&=& (2\pi)^4\delta^4(p_1+p_2-p_3-p_4)\nonumber \\
&\times& iD_A(p_1) iD_B(p_2) iT_{AB,AB}(p_1,p_2;p_3,p_4) iD_A(p_3) iD_B(p_4)
\end{eqnarray}
where $D_{A(B)}(p)$ is the free propagator for a particle $A(B)$ in the momentum space,
which does not contain the negative energy part ({\it i.e.}  the contribution of the corresponding anti-particle).
In the above expression, 
${\cal I}(\bf r)$, which is exponentially suppressed for large $r=\vert\br\vert$ as $e^{-c r}$ with $c \propto \sqrt{E_{th}^2-W^2}> 0$,
represents contributions from other than elastic scattering $A+B\rightarrow A+B$.

With the partial  wave decomposition that
\begin{eqnarray}
\psi_{AB}^W(\br) &=& 4\pi\sum_{l,m} i^l \psi_{AB,l}^W(r,k) Y_{lm}(\Omega_{\brs})\overline{Y_{lm}(\Omega_{\bks})}
\end{eqnarray}
where $k=\vert\bk\vert$, $Y_{lm}$ is the spherical harmonic function  and  $\Omega_{\brs}$ is the solid angle  of the vector $\br$, 
one can show  for the large $r$ that 
\begin{eqnarray}
\psi_{AB,l}^W(r,k) &\rightarrow& A_l \frac{\sin( k r -l\pi/2 +\delta_l(k))}{k r}, 
\end{eqnarray}  
where the "phase shift" $\delta_l(k)$ is the phase of the S-matrix of the $A+B\rightarrow A+B$ scattering for the partial wave $l$. Therefore the NBS wave function is indeed the "wave function"
which describes the  $AB\rightarrow AB$ elastic scattering\cite{Lin:2001ek,Aoki:2005uf,Ishizuka:2009bx,Aoki:2009ji}. 

In the finite volume, restricted values of $k$ denoted  by $k_n$ can be realized to satisfy the boundary condition. From the energy of two particle $W_L$ in the finite volume, one can determine the phase shift $\delta_l (k_n)$ through L\"uscher's formula\cite{Luscher:1990ux}, where
$k_n$ is determined from $W_L=E_{k_n}^A+E_{k_n}^B$.

\subsection{Strategy to define the potential}
In this subsection, we summarize our strategy to define the "potential" in QCD.
\begin{enumerate}
\item[(1)] We choose the field operator $\varphi_A$ and $\varphi_B$. If these operators are composite,
there are many choices to create the same one-particle state. For example, we take the local operator for nucleon in our previous calculations for the nucleon potential\cite{Ishii:2006ec,Aoki:2008hh,Aoki:2009ji}.
\item[(2)] We then measure the NBS wave function, defined by
\begin{eqnarray}
\psi_{AB}^W(\br)  &=& \lim_{\delta\rightarrow 0^+}
\langle 0 \vert  T\{\varphi_{A}(\bx+\br,\delta) \varphi_{B}(\bx,0)\} \vert AB; W,\bP=0\rangle .
\end{eqnarray}   
\item[(3)] Motivated by the fact that the NBS wave function describes the elastic scattering in the large $r$, we 
define the non-local potential as
\begin{eqnarray}
\left[E^{AB}_k-H_0^{AB}\right]\psi_{AB}^W(\bx)  &=& 
\int d^3 y\ U(\bx;\by) \psi_{AB}^W(\by) , \\
  E^{AB}_k&=&\frac{k^2}{2\mu_{AB}},\quad   H_0^{AB} =\frac{-\nabla^2}{2\mu_{AB}},\nonumber
\end{eqnarray} 
where $k=\vert\bk\vert$ and the reduced mass $\mu_{AB}$ is defined by $1/\mu_{AB}=1/m_A + 1/m_B$.
\item[(4)] We then perform the velocity (or derivative) expansion that
$U(\bx;\by) = V(\bx,{\bf\nabla})\delta^3(\bx-\by)$.
In the case of the $NN$ scattering, for example, we have\cite{okubo}
\begin{eqnarray}
V(\br,{\bf\nabla}) = V_0(r) + V_\sigma(r) (\bsigma_1\cdot\bsigma_2) + V_T(r) S_{12} +
V_{\rm LS}(r)\bL\cdot\bS +O({\bf\nabla}^2)
\end{eqnarray}
where $\bsigma_i$ is the spin operator of $i$-th particle, $S_{12}$ is the tensor operator given by
\begin{eqnarray}
S_{12} =\frac{3}{r^2} ({\bsigma_1}\cdot\br)({\bsigma_2}\cdot\br)-({\bsigma_1\cdot\bsigma_2}).
\end{eqnarray}
In the above expansion, the first three terms are of the leading order (LO), which do not contain derivatives, while the 4-th term is of the next leading order (NLO) with one derivative.
\item[(5)] Once we obtain the potentials, we solve the Schr\"odinger equation with these potentials in the {\it infinite} volume  to obtain physical observables such as the phase shift and binding energy.
Note that the exact value of the scattering phase shift $\delta(k)$ is obtained from this Schr\"odinger equation,  while $\delta(k^\prime)$ at $k^\prime\not= k$ is approximated one
as long as the derivative expansion for the potential is truncated at the finite order.
\end{enumerate}

In order to extract the NBS wave function on the lattice, we evaluate the 4-point  correlation function
as
\begin{eqnarray}
{\cal G}_{AB} (\bx,\by,t-t_0;J^P)&\equiv& \langle 0\vert T\{\varphi_A(\bx,t)\varphi_B(\by,t)\}
\overline{\cal J}_{AB}(t_0;J^P)\vert 0 \rangle \\
&=&\sum_n A_n \langle 0\vert T\{\varphi_A(\bx,0)\varphi_B(\by,0)\}\vert W_n\rangle e^{-W_n(t-t_0)} \\
&\rightarrow & A_0\psi_{AB}^{W_0}(\br=\bx-\by;J^P)e^{-W_0(t-t_0)}, \quad t-t_0 \rightarrow\infty
\end{eqnarray}
where $A_n=\langle W_n\vert \overline{\cal J}_{AB}(t_0;J^P)\vert 0\rangle$, and
$\overline{\cal J}_{AB}(t_0;J^P)$ is some source operator which create 2-particle states of $AB$ at $t_0$ with fixed total angular momentum $J$ and parity $P$\cite{Ishii:2006ec,Aoki:2008hh,Aoki:2009ji}.

\subsection{Frequently Asked Questions }
There exist some questions to the definition of the potential in the previous subsection.
\begin{itemize}
\item[(1)] Does the potential depend on the choice of operators $\varphi_A, \varphi_B$ ? \\
Yes. The potential  of course depends on the operators $\varphi_A, \varphi_B$ from which the NBS wave function is defined. The choice of the operators can be regarded as the "scheme" to define the potential, since the potential itself is not a physical observable so that it can be scheme-dependent.
If we calculate the physical observables, however, we obtain the unique result irrespective of the choice for the operators $\varphi_A, \varphi_B$ as long as  $U(\bx;\by)$ is evaluated exactly. 
This is quite analogous to the running coupling, which is of course scheme-dependent.
Physical matrix elements do not depend on the scheme of the running coupling, as long as
they are evaluated exactly.
\item[(2)] Does the potential depend on the total energy at which the NBS wave function is defined ? \\
By construction, $U(\bx;\by)$ is non-local but energy-independent, while
the definition
\begin{eqnarray}
V_{\bks }(\bx ) &\equiv & E^{AB}_k -\frac{
H_0^{AB}\psi_{AB}^W(\bx)}{  \psi_{AB}^W(\bx)} 
\end{eqnarray}
gives a local but energy(momentum)-dependent potential. It is easy to see that  $V_{\bks}(\bx)$ is equivalent to $U(\bx,\by)$ since the number of degrees of freedom of $\bk$ is equal to 
that of $\by$ in the finite volume\cite{Aoki:2009ji}.
From the $\bk$-dependence of the $V_{\bks}(\bx)$,
we therefore  can determine the higher order terms of the derivative expansion 
$V(\bx, {\bf\nabla})$. It turns out, however, that the $\bk$-dependence of $V_{\bks}(\bx)$ for the nucleon potential is very small in quenched QCD between $k\simeq 0$ MeV($E_k^{NN}\simeq 0$ MeV) and $k\simeq 240$ MeV($E_k^{NN}\simeq 45$ MeV) for our scheme\cite{Murano:2011nz,Aoki:2008wy,Murano:2010tc,Murano:2010hh}, where the local composite field of three quarks is used for nucleon operator.  

\end{itemize}

\section{Hadron interactions above inelastic threshold}
\label{sec:inelastic}

\subsection{NBS wave function in inelastic scattering}
We now consider  the scatterings $A+B\rightarrow A+B$ and $A+B\rightarrow C+D$.
We assume that $m_A + m_B < m_C + m_D < W $, where $W=E_k^A + E_k^B$ is the total energy of the system with $E_k^X =\sqrt{m_X^2 +\bk^2}$. In this situation, the QCD eigenstate with the quantum numbers of the $AB$ state  and center of mass energy $W$  is expressed in general as
\begin{eqnarray}
\vert W \rangle &=& c_{AB} \vert AB,W\rangle + c_{CD} \vert CD, W\rangle +\cdots\\
\vert AB, W\rangle &=&  \vert A, \bk \rangle_{\rm in}\otimes \vert B, -\bk \rangle_{\rm in}, \quad
\vert CD, W\rangle =  \vert C, \bq \rangle_{\rm in}\otimes \vert D, -\bq \rangle_{\rm in},
\end{eqnarray}
where $W=E_k^A+E_k^B = E_q^C+E_q^D$.
We define the following NBS wave functions,
\begin{eqnarray}
\psi_{AB}(\br,\bk ) &=& \lim_{\delta\rightarrow 0^+}\langle 0 \vert T\{ \varphi_A(\bx+\br,\delta) \varphi_B(\bx,0) \}\vert W \rangle, \\
\psi_{CD}(\br,\bq ) &=& \lim_{\delta\rightarrow 0^+} \langle 0 \vert T\{ \varphi_C(\bx+\br,\delta) \varphi_D(\bx,0) \}\vert W \rangle, 
\end{eqnarray}
which can be expressed as
\begin{eqnarray}
\psi_{AB}(\br,\bk ) &=& \sqrt{Z_A Z_B}\left[
c_{AB} \left\{ e^{i\bks\cdot\brs} +\int \frac{d^3 p}{(2\pi)^3}\frac{e^{i\bps\cdot\brs}}{\bp^2-\bk^2-i\epsilon}
\frac{E_p^A + E_k^A}{4W E_p^A} T^{AB,AB}(p_A,p_B,k_A,k_B)
\right\}\right.\nonumber \\
&+&\left. c_{CD} \int \frac{d^3 p}{(2\pi)^3}\frac{e^{i\bps\cdot\brs}}{\bp^2-\bk^2-i\epsilon}
\frac{E_p^A + E_k^A}{4W E_p^A} T^{AB,CD}(p_A,p_B,q_C,q_D)
\right] \\
\Psi_{CD}(\br,\bq ) &=& \sqrt{Z_C Z_D}\left[
c_{CD} \left\{ e^{i\bqs\cdot\brs} +\int \frac{d^3 p}{(2\pi)^3}\frac{e^{i\bps\cdot\brs}}{\bp^2-\bq^2-i\epsilon}
\frac{E_p^C + E_q^C}{4W E_p^C} T^{CD,CD}(p_C,p_D,q_C,q_D)
\right\}\right.\nonumber \\
&+&\left. c_{AB} \int \frac{d^3 p}{(2\pi)^3}\frac{e^{i\bps\cdot\brs}}{\bp^2-\bq^2-i\epsilon}
\frac{E_p^C + E_q^C}{4W E_p^C} T^{CD,AB}(p_C,p_D,k_A,k_B)
\right] 
\end{eqnarray}
where $p_A =(E_p^A,\bp)$, $p_B=(W-E_p^A,-\bp)$, $p_C =(E_p^C,\bp)$, $p_D=(W-E_p^C,-\bp)$, $k_A=(E_k^A,\bk)$, $k_B=(E_k^B,-\bk)$, $q_C=(E_q^C,\bq)$ and $q_D=(E_q^D,-\bq)$.

Introducing
\begin{eqnarray}
H^{AB,AB(CD)}(\bp,\bk(\bq)) &=& \frac{E_p^A + E_k^A}{4W E_p^A} T^{AB,AB(CD)}(p_A,p_B,k_A,k_B (q_C,q_D) ) \\
H^{CD,AB(CD)}(\bp,\bk(\bq))&=& \frac{E_p^C + E_q^D}{4W E_p^C} T^{CD,AB(CD)}(p_C,p_D,k_A,k_B(q_C,q_D) )
\end{eqnarray}
and
using the partial wave decomposition such that\footnote{Here we ignore spins for simplicity.},
\begin{eqnarray}
\psi_{XY}(\br,\bk) &=& 4\pi\sum_{l,m} i^l \psi^l_{XY}(r,k)Y_{lm}(\Omega_{\brs}) \overline{Y_{lm}(\Omega_{\bks})} \\
e^{i\bks\cdot\brs}&=& 4\pi \sum_{l,m} i^l j_l(kr) Y_{lm}(\Omega_{\brs}) \overline{Y_{lm}(\Omega_{\bks})} \\
H^{XY,VZ}(\bp,\bk) &=& 4\pi \sum_{l,m} H_l^{XY,VZ} (p,k) Y_{lm}(\Omega_{\bps}) \overline{Y_{lm}(\Omega_{\bks})}
\end{eqnarray}
with $XY,VZ=AB$ or $CD$, we have
\begin{eqnarray}
\psi_{AB}^l(r,k) &=& \sqrt{Z_A Z_B}\left[ c_{AB}\left\{
j_l(kr) + \int \frac{p^2dp}{2\pi^2} \frac{1}{p^2-k^2-i\epsilon} H_l^{AB,AB}(p,k)j_l(pr)\right\}\right.\nonumber\\
&+&\left.
c_{CD} \int \frac{p^2dp}{2\pi^2} \frac{1}{p^2-k^2-i\epsilon} H_l^{AB,CD}(p,q)j_l(pr)\right] \\
\psi_{CD}^l(r,q) &=& \sqrt{Z_C Z_D}\left[ c_{CD}\left\{
j_l(qr) + \int \frac{p^2dp}{2\pi^2} \frac{1}{p^2-q^2-i\epsilon} H_l^{CD,CD}(p,q)j_l(pr)\right\}\right.\nonumber\\
&+&\left.
c_{AB} \int \frac{p^2dp}{2\pi^2} \frac{1}{p^2-q^2-i\epsilon} H_l^{CD,AB}(p,k)j_l(pr)\right] .
\end{eqnarray}
Here the spherical harmonic function $Y_{lm}$ is normalized as
\begin{eqnarray}
\int d\Omega_{\brs}\ \overline{Y_{lm}(\Omega_{\brs})} Y_{l^\prime m^\prime} (\Omega_{\brs}) &=& \delta_{ll^\prime}\delta_{mm^\prime} 
\end{eqnarray}
with the solid angle $\Omega_{\brs}$ of the vector $\br$,  and  the spherical Bessel function 
$j_l(x)$ is given by 
\begin{eqnarray}
j_l(x) &=& (-x)^l\left(\frac{1}{x}\frac{d}{dx}\right)^l \left(\frac{\sin x}{x}\right)
\simeq \frac{\sin(x-l\pi/2)}{x}, \quad x\rightarrow \infty .
\end{eqnarray}

The $p$ integral gives\cite{Lin:2001ek,Aoki:2005uf,Ishizuka:2009bx}
\begin{eqnarray}
\psi_{AB}^l(r,k) &=& \sqrt{Z_A Z_B}\left[ c_{AB}\left\{
j_l(kr) + \frac{k}{4\pi} H_l^{AB,AB}(k,k)\left\{ n_l(kr)+ ij_l(kr)\right\} \right\}\right.
\nonumber\\
&+&\left. c_{CD} \frac{k}{4\pi} H_l^{AB,CD}(k,q)\left\{ n_l(kr)+ ij_l(kr)\right\} \right] 
+{\cal I}_{AB}^l(r)\\
\psi_{CD}^l(r,q) &=& \sqrt{Z_C Z_D}\left[ c_{CD}\left\{
j_l(qr) + \frac{q}{4\pi} H_l^{CD,CD}(q,q)\left\{ n_l(qr)+ ij_l(qr)\right\} \right\}
\right.\nonumber\\
&+&\left.
c_{AB} \frac{q}{4\pi} H_l^{CD,AB}(q,k)\left\{ n_l(qr)+ ij_l(qr)\right\} \right] 
+{\cal I}_{CD}^l(r)
\end{eqnarray}
where ${\cal I}_{XY}^l(r)$, which represents all contributions  except those from the pole at $p=k$ or $p=q$,  is exponentially suppressed for large $r$, and another  spherical Bessel function 
$n_l(x)$ is given by 
\begin{eqnarray}
n_l(x) &=& (-x)^l\left(\frac{1}{x}\frac{d}{dx}\right)^l \left(\frac{\cos x}{x}\right)
\simeq \frac{\cos(x-l\pi/2)}{x}, \quad x\rightarrow \infty .
\end{eqnarray}

The unitarity relation
\begin{eqnarray}
T - T^\dagger &=& i T^\dagger T
\end{eqnarray}
for the on-shell T-matrix $T$ of the 2 channel scattering gives
\begin{eqnarray}
 T_l^{I,J}(W) &=& \frac{8\pi W}{p_I}\left[O(W)
\left(
\begin{array}{cc}
e^{i\delta_l^1(W)}\sin \delta_l^1(W) & 0 \\
0 & e^{i\delta_l^2(W)}\sin \delta_l^2(W) \\
\end{array}
\right) 
O^{-1}(W)\right]^{I,J} \\
O(W) &=& \left(
\begin{array}{cc}
\cos\theta(W) & -\sin\theta(W) \\
\sin\theta(W) & \cos\theta(W) \\
\end{array}
\right) , \qquad I=1,2\quad J=1,2
\end{eqnarray}
where $\delta_l^i(W)$ is the scattering phase shift , whereas $\theta(W)$ is the mixing angle between 1 and 2. Here index 1 represents $AB$ while 2 represents $CD$. 
With this notation $p_1 = k$ and $p_2=q$. We therefore obtain
\begin{eqnarray}
\frac{p_I}{4\pi} H_l^{I,J}(p_I, p_J) &=& \left[O(W)
\left(
\begin{array}{cc}
e^{i\delta_l^1(W)}\sin \delta_l^1(W) & 0 \\
0 & e^{i\delta_l^2(W)}\sin \delta_l^2(W) \\
\end{array}
\right) 
O^{-1}(W)\right]^{I,J} .
\end{eqnarray}

Using these results, the NBS wave functions of the 2 channel system behave for large $r$ as
\begin{eqnarray}
\left(
\begin{array}{l}
\hat\psi_{AB}^l(r,k) \\
\hat\psi_{CD}^l(r,q) \\
\end{array}
\right)
&\simeq&
\left(
\begin{array}{ll}
 j_l(k r) & 0 \\
 0 &  j_l(q r) \\
 \end{array}
 \right)
\left(
\begin{array}{l}
c_{AB} \\
c_{CD}\\
\end{array}
\right) 
+
\left(
\begin{array}{ll}
 n_l(kr)+i j_l(k r) & 0 \\
 0 & n_l(qr)+i  j_l(q r) \\
 \end{array}
 \right)
\nonumber \\
\nonumber \\
&\times&
O(W)
\left(
\begin{array}{cc}
e^{i\delta_l^1(W)}\sin \delta_l^1(W) & 0 \\
0 & e^{i\delta_l^2(W)}\sin \delta_l^2(W) \\
\end{array}
\right) 
O^{-1}(W)
\left(
\begin{array}{l}
c_{AB} \\
c_{CD}\\
\end{array}
\right) 
\end{eqnarray}
where $ \hat \psi_{XY}^l=\psi_{XY}^l/ \sqrt{Z_XZ_Y}$.
This expression shows that the NBS wave function for large $r$ agree with scattering waves described by two scattering phases $\delta_l^i(W)$ ($i=1,2$) and one mixing angle $\theta(W)$.  

\subsection{Coupled channel potentials}
Let us now consider QCD in the finite volume $V$.
In the finite volume, $ \vert AB, W \rangle $ and $ \vert CD, W \rangle $ are no longer eigenstates of the hamiltonian. 
True eigenvalues are shifted from $W$ to $W_i = W + O(V^{-1})$ ($i=1,2$).
By diagonalization method in lattice QCD simulations, 
it is relatively easy to determine $W_1$ and $W_2$. With these values  L\"uscher's finite volume formula gives two conditions, which,
however, are insufficient to determine three observables,
$\delta_l^1$, $\delta_l^2$ and $\theta$. (See \citen{Liu:2005kr, Lage:2009zv,Bernard:2010fp} for recent proposals to overcome this difficulty.)
We here propose alternative approach to extract three observables,
$\delta_l^1$, $\delta_l^2$ and $\theta$, in lattice QCD through
the above NBS wave functions.
We consider the (normalized) NBS wave functions at two different values of energy, $W_1$ and $W_2$,
in the finite volume:
\begin{eqnarray}
\psi_{AB}(\br,\bk_i ) &=& \frac{1}{\sqrt{Z_AZ_B}}\lim_{\delta\rightarrow 0^+}\langle 0 \vert T\{ \varphi_A(\bx+\br,\delta) \varphi_B(\bx,0) \}\vert W_i \rangle \\
\psi_{CD}(\br,\bq_i ) &=& \frac{1}{\sqrt{Z_CZ_D}}\lim_{\delta\rightarrow 0^+} \langle 0 \vert T\{ \varphi_C(\bx+\br,\delta) \varphi_D(\bx,0) \}\vert W_i \rangle, \quad i=1,2 .
\end{eqnarray}
(We here omit $\tilde{}$ on $\psi$.)
We then define the coupled channel non-local potentials from the coupled channel Schr\"odinger equation as
\begin{eqnarray}
\left[E^{AB}_{k_i} - H_0^{AB}\right] \psi_{AB}(\bx,\bk_i) &=&
\int d^3 y\ U_{AB,AB}(\bx;\by)\ \psi_{AB}(\by,\bk_i)+ \int d^3 y\ U_{AB,CD}(\bx;\by)\ \psi_{CD}(\by,\bq_i)\nonumber \\
\\
\left[E^{CD}_{q_i} - H_0^{CD}\right] \psi_{CD}(\bx,\bk_i) &=&
\int d^3 y\ U_{CD,AB}(\bx;\by)\ \psi_{AB}(\by,\bk_i)+ \int d^3 y\ U_{CD,CD}(\bx;\by)\ \psi_{CD}(\by,\bq_i) \nonumber \\
\end{eqnarray}
for $i=1,2$.
As before we introduce the derivative expansion as
\begin{eqnarray}
U_{XY,VZ} (\bx;\by) &=& V_{XY,VZ}(\bx,\nabla)\delta^3(\bx-\by)
= \left[V_{XY,VZ}(\bx) + O(\nabla)\right]\delta^3(\bx-\by)
\end{eqnarray}
and at the leading order of the expansion, we have
\begin{eqnarray}
K_{AB}(\bx,\bk_i)\equiv \left[E^{AB}_{k_i} - H^{AB}_0\right] \psi_{AB}(\bx,\bk_i) &=&
 V_{AB,AB}(\bx)\ \psi_{AB}(\bx,\bk_i)+  V_{AB,CD}(\bx)\ \psi_{CD}(\bx,\bq_i)\nonumber \\
\\
K_{CD}(\bx,\bq_i)\equiv \left[E^{CD}_{q_i} - H^{CD}_0\right] \psi_{CD}(\bx,\bk_i) &=&
V_{CD,AB}(\bx)\ \psi_{AB}(\bx,\bk_i)+ U_{CD,CD}(\bx)\ \psi_{CD}(\bx,\bq_i). \nonumber \\
\end{eqnarray}
These equations for $i=1,2$  can be solved as
\begin{eqnarray}
\left(
\begin{array}{ll}
V_{AB,AB}(\bx) &  V_{AB,CD}(\bx) \\
V_{CD,AB}(\bx) & V_{CD,CD}(\bx)\\
\end{array}
\right) &=&
\left(
\begin{array}{ll}
K_{AB}(\bx,\bk_1) & K_{AB}(\bx,\bk_2)\\
K_{CD}(\bx,\bq_1) & K_{CD}(\bx,\bq_2)\\
\end{array}
\right) \nonumber \\
&\times &
\left(
\begin{array}{ll}
\psi_{AB}(\bx,\bk_1) & \psi_{AB}(\bx,\bk_2) \\
\psi_{CD}(\bx,\bq_1) & \psi_{CD}(\bx,\bq_2) \\
\end{array}
\right)^{-1}. 
\end{eqnarray}

Once we obtain the coupled channel local potentials $V_{XY, VZ}(\bx)$, we solve the coupled channel Scr\"odinger equation in {\it infinite} volume with some appropriate boundary condition such that the incoming wave has a definite $l$ and consists of the $AB$ state only , in order to extract
three observables for each $l$, $\delta_l^1(W)$, $\delta_l^2(W)$ and $\theta(W)$,  
at all values of $W$.
Of course, since $V_{XY,VZ}$ is the leading order approximation in the velocity expansion of $U_{XY,VZ}(\bx;\by)$,  results for three observables $\delta_l^1(W)$, $\delta_l^2(W)$ and $\theta(W)$  at $W\not=W_1, W_2$ are also approximate ones and might be different from the exact values.  By performing an  additional extraction of $V_{XY, VZ}(\bx)$ at $(W_3,W_4)\not=( W_1,W_2)$, we can test  how good the leading order approximation is, as in the case of the elastic scattering\cite{Murano:2011nz}.

The above result can be easily extended to the coupled channel among $n$ states $ A_I B_I$ ($I=1,2,\cdots, n$) .
Local potentials at the LO in the velocity expansion are given by
\begin{eqnarray}
V_{I,J}(\br) &=& \sum_{M=1}^n K_I(\br, \bk^I_M) X^{-1}(\br)_{MJ} \\
K_I(\br, \bk^I_M) &\equiv& \left[E^I_M-H^I_0\right]  \psi_I(\br,\bk^I_M)
\end{eqnarray}
where $X^{-1}(\br)$ is the inverse of the $n\times n$ NBS wave function matrix 
\begin{eqnarray}
X(\br)_{IJ} &\equiv& \psi_I(\br,\bk^I_J)
= \frac{1}{\sqrt{Z_{A_I}Z_{B_I}}}\lim_{\delta\rightarrow 0^+}\langle 0 \vert T\{ \varphi_{A_I}(\bx+\br,\delta) \varphi_{B_I}(\bx,0) \}\vert W_J \rangle,
\end{eqnarray}
the momentum $\bk^I_J$ satisfies  the relation that
$W_J=\sqrt{(\bk^I_J)^2+m_{A_I}^2}+\sqrt{(\bk^I_J)^2+m_{B_I}^2}$ for $J=1,2,\cdots, n$, 
$E^I_J = (\bk^I_J)^2/(2\mu_{A_IB_I})$ and $H^I_0 = -\nabla^2/(2\mu_{A_IB_I})$.

\subsection{Inelasticity and non-locality}
We now consider the 2-channel problem again.
Let us assume that the leading order approximation for the coupled channel potential works reasonably well. Instead of considering the coupled channel potential, the effective potential for the $AB$ channel is given by 
\begin{eqnarray}
U_{AB,AB}^{\rm eff}(\bx,\by) &=& V_{AB,AB}(\bx)\delta^{(3)}(\bx-\by) \nonumber \\
&+& V_{AB,CD}(\bx)\frac{1}{E^{CD}_{q}-H_0^{CD}-V_{CD.CD}}(\bx,\by) V_{CD,AB}(\by), 
\label{eq:eff}
\end{eqnarray}
where the non-locality becomes manifest in the second term. The magnitude of momentum $\bq$ for the $CD$ channel is expressed as
\begin{eqnarray}
4W^2 \bq^2 &=& (W^2-(m_C+m_D)^2 )(W^2-(m_C-m_D)^2 ).
\end{eqnarray}
To estimate the magnitude of non-locality in eq.(\ref{eq:eff}), we here ignore the $V_{CD,CD}$ term in the denominator. In this case, if the total energy $W$ is below the inelastic threshold $m_C+m_D$ such that $\bq^2= -M^2 < 0$, we have
\begin{eqnarray}
\frac{1}{H_0^{CD}-E^{CD}_{q}}(\bx,\by)
= \frac{2\mu_{CD}}{4\pi \vert\bx-\by\vert} e^{-M\vert\bx-\by\vert} 
\label{eq:inequality}
\end{eqnarray}
Therefore, non-locality of the potential $U^{\rm eff}_{AB,AB}(\bx,\by)$ is exponentially suppressed as long as $M$ is large enough.
As $W$ approaches $m_C+m_D$, however, non-locality of the potential becomes larger and manifest.
In this simple example, this non-locality of the single channel potential can be completely removed by introducing the $CD$ state and  {\it coupled channel} potentials between $AB$ and $CD$. 
Therefore, in practice, the non-locality caused by inelastic final states whose thresholds are closed to the total energy $W$ is expected to become milder  for  the  coupled channel potentials including these states.

\section{Extension: Inelastic scattering with particle production}
\label{sec:extension}

The method considered in the previous section can be generalized to inelastic scattering where a number of particles is not conserved. For illustration, we consider  a case that the scatterings $A+B\rightarrow A+B$ and $A+B\rightarrow A+B+C$ occur and the total energy $W$ satisfies 
$ m_A+m_B+m_C < W < m_A+m_B+2m_C$.

We consider the following NBS wave functions in the center of mass system:
\begin{eqnarray}
\psi_{AB}^W(\bx) &=& \frac{1}{\sqrt{Z_AZ_B}}\lim_{\delta\rightarrow 0^+}\langle 0 \vert \varphi_A(\br+\bx,\delta)\varphi_B(\br,0)\vert W \rangle \\
\psi_{ABC}^W(\bx,\by) &=& \frac{1}{\sqrt{Z_AZ_BZ_C}}\lim_{\delta\rightarrow 0^+}\langle 0 \vert \varphi_A(\br+\bx+\frac{\by\,\mu_{BC}}{m_C},\delta)\varphi_B(\br+\by,0)\varphi_C(\br,-\delta)\vert W \rangle ,
\end{eqnarray}
where
\begin{eqnarray}
\vert W \rangle &=& c_1\ \vert \bk \rangle_{\rm in}\otimes \vert -\bk \rangle_{\rm in}  +
c_2\ \vert \bq_x \rangle_{\rm in}\otimes \vert \bq_y-\bq_x\frac{\mu_{BC}}{m_C} \rangle_{\rm in}\otimes \vert -\bq_y- \bq_x\frac{\mu_{BC}}{m_B}\rangle_{\rm in}
\end{eqnarray}
with
\begin{eqnarray}
W &=& \sqrt{\bk^2+m_A^2} +\sqrt{\bk^2+m_B^2} \nonumber \\
&=& \sqrt{\bq_x^2+m_A^2}+\sqrt{(\bq_y-\bq_x\frac{\mu_{BC}}{m_C})^2+m_B^2} +\sqrt{(\bq_y+
\bq_x\frac{mu_{BC}}{m_B})^2+m_C^2} 
\end{eqnarray}
and $1/\mu_{BC} =1/m_B + 1/m_C$. Here  $\by = \br_B-\br_C$ is a relative coordinate between 
$B$ and $C$ with the reduced mass $\mu_{BC}$, while $\bx =\br_A-{\bf R}_{BC}$ is the one between $A$ and the center of mass of $B$ and $C$ with ${\bf R}_{BC} = (m_B\br_B + m_C \br_C)/(m_B+ m_C)$.  We here assume the property that $\psi_{ABC}^W$ has asymptotic behavior of the scattering wave of $A+B+C$ as $\vert \bx\vert, \vert \by\vert \rightarrow\infty$, as shown for $\psi_{AB}^W$ in the previous section.  Although this property has not been shown so far, it is reasonable to assume this.
We leave the proof for this important property to the future investigation.  

We define the non-local potential from the coupled channel equations as
\begin{eqnarray}
K_{AB}^W(\bx)&\equiv& \left[E_k^{AB}-H_{0,x}^{AB}\right] \psi_{AB}^W(\bx)= 
\int d^3\,z\ U_{AB,AB}(\bx;\bz)\, \psi_{AB}^W(\bz) \nonumber \\
&+& \int d^3\,z\ d^3\,w\ U_{AB,ABC}(\bx;\bz,\bw)\, \psi_{ABC}^W(\bz,\bw)\\
K_{ABC}^W(\bx,\by)&\equiv&\left[E_{q_x}^{A,BC}+E_{q_y}^{BC} -H_{0,x}^{A,BC} - H_{0,y}^{BC}\right]
\psi_{ABC}^W(\bx,\by)= 
\int d^3\,z\ U_{ABC,AB}(\bx,\by;\bz)\nonumber \\
&\times& \psi_{AB}^W(\bz)\nonumber 
+ \int d^3\,z\ d^3\,w\ U_{ABC,ABC}(\bx,\by;\bz,\bw)\, \psi_{ABC}^W(\bz,\bw)
\end{eqnarray}
where
\begin{eqnarray}
H_{0,x}^{AB}&=&\frac{-\nabla_{\bxs}^2}{2\mu_{AB}}, \quad
H_{0,x}^{A,BC}=\frac{-\nabla_{\bxs}^2}{2\mu_{A,BC}}, \quad
H_{0,y}^{BC}=\frac{-\nabla_{\bys}^2}{2\mu_{BC}}, \\
E_k^{AB} &=& \frac{\bk^2}{2\mu_{AB}},\quad
E_{q_x}^{A,BC} = \frac{\bq_x^2}{2\mu_{A,BC}} ,\quad
E_{q_y}^{BC} = \frac{\bq_y^2}{2\mu_{BC}} 
\end{eqnarray}
with another reduced mass defined by $1/\mu_{A,BC}=1/m_A+1/(m_B+m_C)$.

We consider the following velocity expansions
\begin{eqnarray}
U_{AB,AB} (\bx; \bz) &=& \left[V_{AB,AB}(\bx) + O(\nabla_x)\right]\delta^3(\bx-\bz) \\
U_{AB,ABC} (\bx; \bz,\bw) &=& \left[V_{AB,ABC}(\bx,\bw) + O(\nabla_x)\right]\delta^3(\bx-\bz)  \\
U_{ABC,AB} (\bx,\by; \bz) &=& \left[V_{ABC,AB}(\bx,\by) + O(\nabla_x)\right]\delta^3(\bx-\bz)\\
U_{ABC,ABC} (\bx,\by; \bz,\bw) &=& \left[V_{ABC,ABC}(\bx,\by) + O(\nabla_x,\nabla_y)\right]\delta^3(\bx-\bz) \delta^3(\by-\bw) ,
\end{eqnarray}
where the hermiticity of the non-local potentials gives $V_{AB,ABC}(\bx,\by) = V_{ABC,AB}(\bx,\by)$.

At the leading order of the velocity expansions, the coupled channel equations become
\begin{eqnarray}
K_{AB}^W(\bx) &=& V_{AB,AB}(\bx) \psi_{AB}^W(\bx) + \int d^3\,w\ V_{AB,ABC}(\bx,\bw)\psi_{ABC}^W(\bx,\bw ) \\
K_{ABC}(\bx,\by) &=& V_{ABC,AB}(\bx,\by)\psi_{AB}^W(\bx) + V_{ABC,ABC}(\bx,\by)\psi_{ABC}^W(\bx,\by) .
\end{eqnarray}
By considering two values of energy such that $W=W_1, W_2$, 
we can determine $V_{ABC,AB}$ and $V_{ABC,ABC}$ from the second equation as
\begin{eqnarray}
\left(
\begin{array}{ll}
V_{ABC,AB}(\bx,\by) & V_{ABC,ABC}(\bx,\by) \\
\end{array}
\right) &=& 
\left(
\begin{array}{ll}
K_{ABC}^{W_1}(\bx,\by) & K_{ABC}^{W_2}(\bx,\by) \\
\end{array}
\right) \nonumber \\
&\times&
\left(
\begin{array}{ll}
\psi_{AB}^{W_1}(\bx) & \psi_{AB}^{W_2}(\bx) \\
\psi_{ABC}^{W_1}(\bx,\by) & \psi_{ABC}^{W_2}(\bx,\by) \\
\end{array}
\right)^{-1} .
\end{eqnarray}
Using the hermiticity  $V_{AB,ABC}(\bx,\by) = V_{ABC,AB}(\bx,\by)$, we can extract $V_{AB,AB}$ from the first equation as
\begin{eqnarray}
V_{AB,AB}(\bx) &=&\frac{1}{\psi_{AB}^W(\bx)}\left[
K_{AB}^{W}(\bx) -\int d^3\,w\ V_{ABC,AB}(\bx,\bw)\psi_{ABC}^W(\bx,\bw ) 
\right]
\end{eqnarray}
for $W=W_1, W_2$. A difference of $V_{AB,AB}(\bx)$ between two estimates at $W_1$ and $W_2$ gives an estimate for higher order contributions in the velocity expansions. 

Once we obtain $V_{AB,AB}$, $V_{AB,ABC}=V_{ABC,AB}$ and $V_{ABC,ABC}$, we can solve the coupled channel Schr\"odinger equations in the {\it infinite} volume, in order to extract physical observables. 
 As $W$ increases and becomes larger than $m_A+m_B + n m_C$, the inelastic scattering  $A+B\rightarrow A+B+n C $ becomes possible. As in the case of $A+B\rightarrow A+B+C$ in the above,  we can define the coupled channel potentials including this channel, though calculations of the NBS wave functions for multi-hadron operators become more and more difficult in practice.  

\section{Summary}
\label{sec:summary}
In this paper, we propose extensions of the method of extracting potentials through NBS wave functions to the case where inelastic scatterings becomes important. We first consider the case that $A+B\rightarrow C+D$ scattering occurs and present an explicit formula to extract the coupled channel potentials  between $AB$ and $CD$. The general formula for the scattering among  $n$ states is also given.  An extension to the case where the particle production occurs during the scattering such as $A+B\rightarrow A+B+C$ is also considered. This can also be extended to more general cases such as  $A+B\rightarrow A+B+nC$ ($n=1,2,3,\cdots .$).

Recently the potential method has been used to extract potentials for the flavor SU(3) limit in lattice QCD where up, down and strange quark masses are all equal\cite{Inoue:2010hs}. In this limit there exist 6-independent potentials corresponding to the irreducible representations of the flavor SU(3),  and  among these, the flavor singlet potential  is strong attractive, suggesting an existence of a bound state in this channel. This expectation has been confirmed by more detailed study in lattice QCD and a bound state of two up quarks, two down quarks and two strange quarks, called the $H$ dibaryon, indeed exists in the flavor SU(3) limit\cite{Inoue:2010es}.
To study the property of the $H$ dibaryon in the real world where the strange quark is much heavier than up and down quarks so that the flavor SU(3) symmetry is broken,  the extension presented in this paper for the coupled channels  is indeed necessary:
The SU(3) breaking in nature appears in the octet baryon mass as $m_N = 939$ MeV, $m_\Lambda = 1116$MeV, $m_\Sigma = 1193$ MeV and $m_\Xi =1318$ MeV.
Therefore  thresholds of two baryon systems with strangeness $S=-2$ and ispspin $I=0$,
to which the $H$ dibaryon belongs, are given by
\begin{eqnarray}
W_{\Lambda\Lambda} = 2232\, \mbox{MeV} < W_{N\Xi} = 2257\, \mbox{MeV} < 
W_{\Sigma\Sigma} = 2386\, \mbox{MeV}.
\end{eqnarray}
A study for coupled channel potentials in this case has started  using the 2+1 flavor lattice QCD and a preliminary result  has already been presented\cite{Sasaki:2010bh}. More details results for this study will be  published soon.


\section*{Acknowledgements}

The authors thank Dr. N. Ishizuka  
for useful discussions.
This research  was supported  in part by the
Grant-in-Aid of MEXT (Nos. 15540254, 18540253, 20340047) and by
Grant-in-Aid for Scientific Research on Innovative Areas (No. 2004:  20105001, 20105003).

\end{document}